\documentclass[aps,twocolumn,floatfix,superscriptaddress,preprintnumbers,showpacs,showkeys,nofootinbib]{revtex4}

\usepackage{amssymb}
\usepackage{amsmath}
\usepackage{amsfonts}
\usepackage{graphicx}
\usepackage{color}
\usepackage{xspace}
\usepackage{ulem}
\usepackage{lscape}
\usepackage{ wasysym }
\usepackage[toc,page]{appendix}
\usepackage{pdfpages}
\usepackage{multirow,rotating}
\usepackage{lipsum}

\def\gsim{\raise0.3ex\hbox{$\;>$\kern-0.75em\raise-1.1ex\hbox{$\sim\;$}}}
\def\lsim{\raise0.3ex\hbox{$\;<$\kern-0.75em\raise-1.1ex\hbox{$\sim\;$}}}

\def\znbb{0\nu\beta\beta}
\def\meff{\langle m_{\nu} \rangle}

\def\qnu{q_{\nu} \hspace{-0.9em}/}

\newcommand{\ba}[1]{\begin{eqnarray} \label{(#1)}}
\newcommand{\ea}{\end{eqnarray}}

\newcommand{\AddrAHEP}{
  {\it AHEP Group, Instituto de F\'{\i}sica Corpuscular --
    CSIC./Universitat de Val{\`e}ncia \\
    Edificio de Institutos de Paterna, Apartado 22085,
  E--46071 Val{\`e}ncia, Spain}}

\newcommand{\AddrUFSM}{
Universidad T\'ecnica Federico Santa Mar\'\i a, \\ 
Centro-Cient\'\i fico-Tecnol\'{o}gico de Valpara\'\i so, \\ 
Casilla 110-V, Valpara\'\i so,  Chile}

\def\gsim{\raise0.3ex\hbox{$\;>$\kern-0.75em\raise-1.1ex\hbox{$\sim\;$}}}
\def\lsim{\raise0.3ex\hbox{$\;<$\kern-0.75em\raise-1.1ex\hbox{$\sim\;$}}}

\begin{document}

\preprint{IFIC/16-70}  
\title{QCD corrections and long-range mechanisms of neutrinoless
  double beta decay}

\author{C. Arbel\'aez}\email{carolina.arbelaez@usm.cl}\affiliation{\AddrUFSM}
\author{M. Gonz\'alez} \email{marcela.gonzalezp@usm.cl}\affiliation{\AddrUFSM}
\author{M. Hirsch} \email{mahirsch@ific.uv.es}\affiliation{\AddrAHEP}
\author{S.G. Kovalenko}\email{Sergey.Kovalenko@usm.cl}\affiliation{\AddrUFSM}



\begin{abstract}
Recently it has been demonstrated that QCD corrections are numerically
important for short-range mechanisms (SRM) of neutrinoless double beta
decay ($\znbb$) mediated by heavy particle exchange. This is due to
the effect of color mismatch for certain effective operators, which
leads to mixing between different operators with vastly different
nuclear matrix elements (NMEs).  In this note we analyze the QCD
corrections for long-range mechanisms (LRM), due to diagrams with
light-neutrino exchange between a Standard Model (V-A)$\times$(V-A)
and a beyond the SM lepton number violating vertex.  We argue that in
contrast to the SRM in the LRM case, there is no operator mixing from
color-mismatched operators.  This is due to a combined effect of the
nuclear short-range correlations and color invariance. As a result,
the QCD corrections to the LRM amount to an effect no more than
60$\%$, depending on the operator in question. Although less crucial,
taken into account QCD running makes theoretical predictions for
$\znbb$-decay more robust also for LRM diagrams.  We derive the
current experimental constraints on the Wilson coefficients for all
LRM effective operators.

\end{abstract}

\maketitle


\section{Introduction}
\label{sec:introduction}

Neutrinoless double beta decay ($\znbb$), being a lepton number
violating (LNV) process, offers an opportunity to probe physics beyond
the SM in a way complementary or maybe even unavailable for collider
experiments. Great efforts have been made in both theoretical and
experimental work on $\znbb$-decay (for recent reviews see for example
Refs. ~\cite{Deppisch:2012nb,Rodejohann:2011mu}).  For the extraction
of limits on any beyond the SM it is indispensable to have a
reliable theoretical description of all the structural levels involved
in this process: From the underlying LNV process at some supposedly
larger energy scale, through the hadronization to nucleon bound
states, to a reliable description at the nuclear level level, where
double beta decay finally takes place.

\begin{figure}[t]

\includegraphics[width=0.4\linewidth]{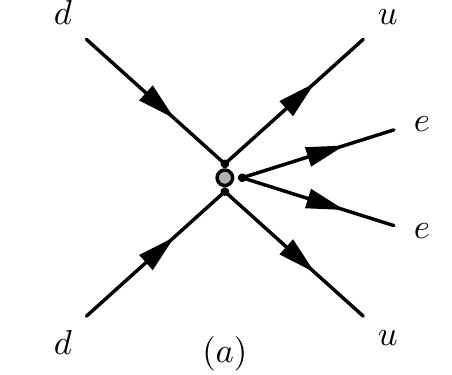}
\includegraphics[width=0.4\linewidth]{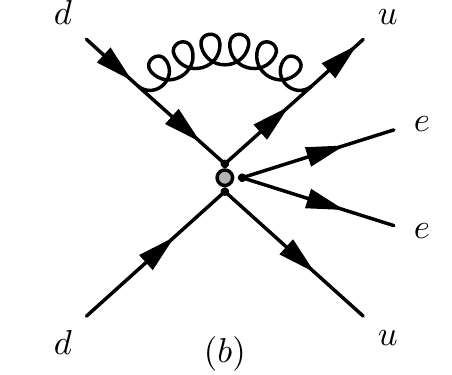}
\includegraphics[width=0.4\linewidth]{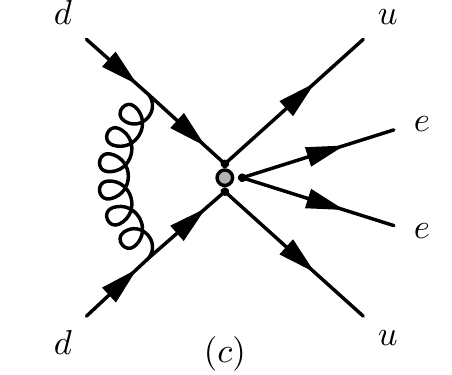}
\includegraphics[width=0.4\linewidth]{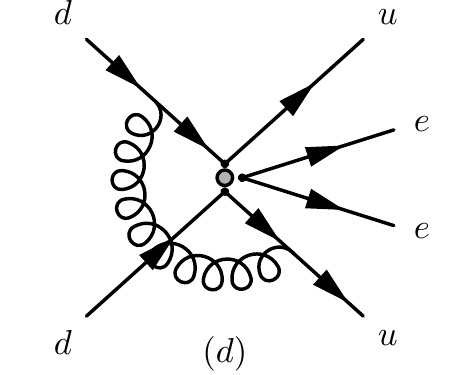}

\caption{Effective $d=9$ operator description of the short-range
  Mechanisms (SRM) of $\znbb$ decay. Diagram (a) gives the tree-level
  description, diagrams (b)-(d) are one-loop QCD corrections to the
  SRM $0\nu\beta\beta$ decay in the effective theory.}
\label{fig:short}
\end{figure}

\begin{figure}[t]
\centering
\includegraphics[width=0.45\linewidth]{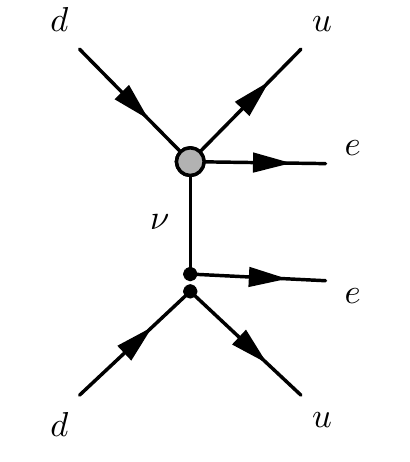}
\includegraphics[width=0.45\linewidth]{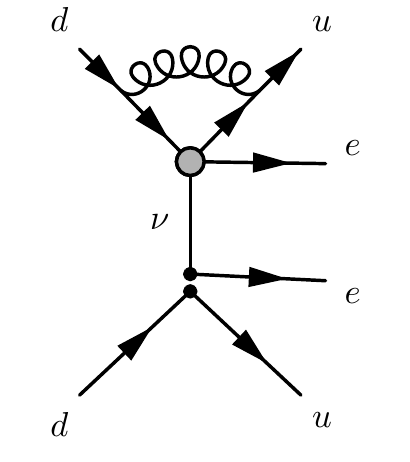}
\caption{Long-range mechanism (LRM) contribution to $\znbb$ decay. The
  diagram on the left shows the tree-level effective diagram: Exchange
  of a light neutrino between a SM charged current vertex and a
  beyond-SM LNV vertex (indicated by the grey blob). The diagram to
  the right shows the one-loop QCD correction to this diagram,
  correcting the right-handed non-SM vertex proportional to $C(\mu)$,
  see text.}
\label{fig:long}
\end{figure}

All these stages have been under scrutiny in the past decades, but it
has now turned out that an important intermediate stage happening
before the hadronization -- namely, QCD-corrections -- has been
overlooked until quite recent works
\cite{Mahajan:2014twa,Gonzalez:2015ady}. In \cite{Gonzalez:2015ady} it
has been shown that the effects of this QCD running can amount to
changes up to $3$ orders of magnitude in the matrix elements of
certain effective dimension-9 operators, describing the short-range
mechanism (SRM) of $\znbb$-decay. Recall that the SRM is a class of
mechanisms mediated by heavy particle exchange as shown in Fig.~\ref{fig:short}.  \ \\ 
Given this surprising result, naturally there appears the question
\cite{Mahajan:2015wpk} if the QCD corrections are also so crucial for
other classes of contributions to $\znbb$ decay, namely, those known
as long-range mechanisms (LRM). LRM are induced by diagrams with
light-neutrino exchange between a Standard Model (SM)
\mbox{(V-A)$\times$(V-A)} and a beyond the SM lepton number violating
(LNV) vertex as shown in Fig.~\ref{fig:long}. In this paper we
analyze the QCD corrections to the LRM and argue that they are
significantly smaller in comparison with the short-range mechanism (SRM) case. We calculate
the RGE improved QCD running of all Wilson coefficients contributing to
LRM and derive the current experimental constraints on these. We 
find that the maximal impact of the QCD running is of the order 
of 60 \%.


\section{Low-energy description of $\znbb$-decay}

Double beta decay takes place at energies much lower than the
electroweak scale. An effective operator description of the process
is therefore adequate.  Let us start by recalling the basic
definitions of the SRM and LRM of $\znbb$-decay. \\
{\it Short-range Mechanisms (SRM) of $\znbb$-decay} encompass all 
high-scale models (HSM) contributing via heavy particle exchange as
in Fig.~\ref{fig:short} with the typical mass $M_{I}$. After
integrating out these heavy degrees of freedom at an energy scale
$\mu< M_{I}$ the SRM is described by the effective Lagrangian
\cite{Pas:2000vn,Gonzalez:2015ady},
\begin{eqnarray}\label{eq:LagGen}
{\cal L}^{0\nu\beta\beta}_{\rm eff} = \frac{G_F^2}{2 m_p} \,
              \sum_{i, XY} C_{i}^{XY}(\mu)\cdot \mathcal{O}^{(9) XY}_{i}(\mu),
\end{eqnarray}
with the complete set of dimension-9 $\znbb$-operators
\cite{Gonzalez:2015ady}
\begin{eqnarray}
\label{eq:OperBasis-1}
\mathcal{O}^{(9) XY}_{1}&=& 4 ({\bar u}P_{X}d) ({\bar u}P_{Y}d) \ j,\\
\label{eq:OperBasis-2}
\mathcal{O}^{(9) XX}_{2}&=& 4 ({\bar u}\sigma^{\mu\nu}P_{X}d)
                         ({\bar u}\sigma_{\mu\nu}P_{X}d) \ j,\\
\label{eq:OperBasis-3}
\mathcal{O}^{(9) XY}_{3}&=& 4 ({\bar u}\gamma^{\mu}P_{X}d) 
                        ({\bar u}\gamma_{\mu}P_{Y}d) \  j,\\
\label{eq:OperBasis-4}
\mathcal{O}^{(9) XY}_{4}&=& 4 ({\bar u}\gamma^{\mu}P_{X}d) 
                         ({\bar u}\sigma_{\mu\nu}P_{Y}d) \ j^{\nu},\\
\label{eq:OperBasis-5}
\mathcal{O}^{(9) XY}_{5}&=& 4 ({\bar u}\gamma^{\mu}P_{X}d) ({\bar u}P_{Y}d) \ j_{\mu},
\end{eqnarray}
where $X,Y = L,R$ and the LNV leptonic currents are 
\begin{eqnarray}\label{eq:Curr}
j = {\bar e}(1\pm \gamma_{5})e^c \, , \quad j_{\mu} = {\bar e}\gamma_{\mu}\gamma_{5} e^c .
\end{eqnarray}
Graphically at low energies $\mu<M_{I}$ the SRM is given by the
pointlike vertex in Fig.~\ref{fig:short} (without gluon lines)
corresponding to the above listed effective operators. The
$C_{i}^{XY}$ in Eq. (\ref{eq:LagGen}) are the Wilson coefficients.
The Wilson coefficients $C_{i}^{XY}$ can be expressed in terms of the
parameters of a particular HSM at a scale $\Lambda \sim M_{I}$, called
``matching scale''.  

{\it Long-range mechanisms (LRM) of $\znbb$-decay} originate from some
HSM with LNV interactions involving heavy particles of a mass $M_{I}$
{\em and a light neutrino $\nu$} as in the top vertex of the diagrams in
Fig.~\ref{fig:long}. The SM charged current interaction of the
neutrino in the bottom vertex completes the $\znbb$-decay diagram.  In
the low energy limit, at scales $\mu< M_{I}$, the heavy block can be
represented by the following complete set of dimension-6 effective LNV
operators:
\begin{eqnarray}
\label{eq:hc3-1}
\mathcal{O}_{1}^{(6) X} &=& 4 (\bar{u} P_{X} d)  \left(\bar{e} P_{R} \nu^{C}\right), \\ 
\label{eq:hc3-2}
\mathcal{O}_{2}^{(6) X} &=& 4  (\bar{u} \sigma^{\mu\nu}P_{X} d) \left(\bar{e}\sigma^{\mu\nu}  P_{R} \nu^{C}\right),\\
\label{eq:hc3-3}
\mathcal{O}_{3}^{(6) X} &=& 4  (\bar{u} \gamma_{\mu}P_{X} d) 
\left(\bar{e} \gamma^{\mu} P_{R} \nu^{C}\right)
\end{eqnarray}
with $X= R, L$.  Then $\znbb$-decay is described by second-order
perturbation theory in the effective Lagrangian \cite{Pas:1999fc}:
\begin{equation}
\mathcal{L}_{\rm eff}^{d=6}=\frac{G_F}{\sqrt{2}} \left( j^{\mu}J^{\dag}_{\mu}+  
\sum_{i}C^{X}_{i}(\mu)\mathcal{O}^{(6)X}_{i}(\mu)\right). 
\label{eq:lrlagrangian}
\end{equation}
Here the first term is the SM low-energy 4-fermion effective
interaction of the currents
\begin{eqnarray}\label{eq:V-A}
&&j^{\mu} = \bar{e}\gamma^{\mu}(1 - \gamma_{5}) \nu, \ \ \ 
J_{\mu} = \bar{d}\gamma_{\mu}(1 - \gamma_{5}) u.
\end{eqnarray} 
We consider only LNV $\Delta L =2$ effective operators so that the LNV
part of the neutrino propagator, proportional to the light neutrino
Majorana mass $\meff$, does not contribute.  This fact is reflected in
the chirality structure of these diagrams $P_{L} (\qnu + m_{\nu})
P_{R}$. Thus we deal with the momentum dependent LRM of $\znbb$-decay,
${\cal A} \propto \qnu$. Contributions that we neglect are 
proportional to $\meff\cdot C_{k}$, i.e. vanish in the limit 
$\meff \to 0$. 

\section{Differences in the RGE evolution of SRM and LRM.}

Both the operators $\mathcal{O}_{i}(\mu)$ and their Wilson
coefficients $C_{i}(\mu)$ in Eqs.~(\ref{eq:LagGen}) 
(\ref{eq:lrlagrangian}) depend on the energy scale $\mu \leq \Lambda$
due to the effect of the QCD loop corrections shown in
Figs.~\ref{fig:short}, \ref{fig:long}. At the ``matching'' scale
the $C_{i}(\mu)$ are calculated in terms of underlying HSM parameters,
like heavy masses and couplings and then QCD-run down to a scale $\mu
= \mu_{0}$, close to the typical $\znbb$-scale.

Although the QCD running is only logarithmic, in some specific cases
mixing of different operators can occur.  Because of the vast
difference of the nuclear matrix elements (NMEs) of some operators, this effect can have a
dramatic impact on the prediction for some particular HSM contributing
to $\znbb$-decay.  This happens, as shown recently in
Ref.~\cite{Gonzalez:2015ady}, in the case of the SRMs, where the
effect may reach $3$ orders of magnitude at the level of amplitude.
Here, we discuss that operator mixing is not important for the case of
the LRM of $\znbb$-decay, We arrive at this conclusion analyzing
analogously both mechanisms, SRM and LRM, passing from the elementary
quark-level $\Delta L = 2$ processes $d d \rightarrow u u + 2 e^{-}$
(SRM) and $d \rightarrow u + e^{-} \nu$ (LRM) to the hadronic level
process $n\, n\rightarrow p\, p + 2 e^{-}$ taking place inside a
$\znbb$-decaying nucleus. One can distinguish the following stages for
the $\znbb$-transition at the different structural levels.

{\it For the SRM} depicted in Fig.~\ref{fig:short}:

{\bf (i)} Two colorless objects -- initial neutrons -- need to approach
each other very closely and form a colorless six-quark $(uudddd)$
state. Note that this configuration is heavily suppressed by the well-known
nuclear effect of ``short-range correlations'' due to the repulsive
nuclear hard core;

{\bf (ii)} Within this six-quark-state occurs the transition \mbox{$d\,
  d\rightarrow u\, u +2e^{-}$} induced by a pointlike QCD-singlet
vertex operator , induced by one of the operators in
Eqs.~(\ref{eq:OperBasis-1})-(\ref{eq:OperBasis-5});
\footnote{There exists another modality of this mechanism not
  requiring the stage (i), instead neutrons emit virtual pions and
  (ii) is realized in $\pi^{-}\pi^{-} \rightarrow 2 e^{-}$. This
  pion-mechanism \cite{Faessler:1996ph,Faessler:1998qv}, less
  suppressed by the short-range correlation, requires a special study. 
  For QCD corrections for the pion mechanism see \cite{Peng:2015haa}.}

{\bf (iii)} At this stage the QCD corrections in Fig.~\ref{fig:short}
have to be considered. In the diagrams in Figs.~\ref{fig:short}(c,d)
the gluon links the quarks from the different color-singlet currents
leading to color-mismatched operators in the final state. The Fiertz
rearrangement of the quark fields in the QCD-corrected operators to
new color-singlet combinations generate operator structures distinct
in some cases from the original bare one;

{\bf (iv)} Finally, a new color-singlet $(uuuudd)$-state projects
onto the final $pp$-state.

{\it For the LRM} shown in Fig.~\ref{fig:long} the situation is
essentially different. The fundamental pointlike $\Delta L =2$
interaction $d\rightarrow u+ e^{-} + \nu$ takes place inside the
nucleon leading to the nucleon-level transition $n\rightarrow p +
e^{-} +\nu$ with the virtual neutrino initiating the $\znbb$-decay
as shown in Fig.~\ref{fig:long}. Nucleons in this case interact at a
distance larger than the repulsive nucleon hard core. This distance is
controlled by the neutrino potential. The average value of the
momentum $q_{\nu}$ flowing in the neutrino propagator is about
$\langle q_{\nu}\rangle \sim p_{F}\sim 100-200$ MeV. Then, reasoning
schematically,
\begin{eqnarray}\label{eq:Majagan-1}
(\bar{u}\Gamma_{i} d) \frac{\qnu}{q_{\nu}^{2}} (\bar{u}\gamma P_{L} d) \rightarrow  
\frac{1}{\langle q_{\nu}\rangle}(\bar{u}\Gamma_{i} d)  (\bar{u}\gamma P_{L} d)
\end{eqnarray}
one may wish to approximate the underlying process by a process which,
formally, looks like a pointlike interaction in the rhs.  It is then
tempting to think \cite{Mahajan:2015wpk} that, as in the case of the
SRM, there are diagrams as in Figs.~\ref{fig:short}(c,d) linking
different color singlet currents.  However, in fact the two initial
d-quarks are located in the two separate initial neutrons and these
are separated by a distance $d \sim \langle q_{\nu}\rangle^{-1}$ (which
is larger than the hard core).

Thus, a gluon exchange between two color singlet nucleons would give
rise to a color nonsinglet final state at the hadronic level.
However, the final hadronic state must be a color singlet to have a
nonzero projection on the nucleon state including two protons. Then 
another gluon exchange in the final state become necessary, which
results in an extra $\alpha_{s}$ suppression. 

Trying to circumvent this issue by putting the two initial neutrons
sufficiently close together, to form a colorless $(uudddd)$ state, is
suppressed by the nuclear hard core.  Thus, for any mechanism with the
quark-level subprocess $d\rightarrow u+ e^{-} + \nu$ we have to deal
with a true long-range $n\, n\rightarrow p\, p + 2 e^{-}$ process
mediated by the exchange with the light neutrino between {\em
  different, distant} nucleons.  Then the only QCD correction to the
color-singlet vertices that should be considered is the one shown in
Fig.~\ref{fig:long}(b). As a result the effect of the QCD running in
the LRM case is not as significant as in the case of SRM.

\section{Calculation of QCD improved Wilson coefficients}

Let us now move on to estimate this effect for LRM numerically.  Let
us first note that we do not care for the QCD corrections to the SM
effective vertex in the bottom of this diagram since: (i) We use the
experimental value of the Fermi constant $G_{F}$ measured at
$\mu_{0}$; (ii) we are only interested in the relation between the
parameters of the $\Delta L =2$ HSMs defined at $\mu=\Lambda$ and the
$\znbb$-decay parameters measured at $\mu = \mu_{0}$.

The QCD corrections to the quark-lepton vertex $V_{QL}$ in the diagram
Fig.~\ref{fig:long} can be written in the general form as:
 \begin{eqnarray}
(\delta V_{QM})^{QCD} &\propto&
 (\bar{u}\gamma^{\nu}\gamma^{\sigma}\Gamma^{NSM}\gamma_{\sigma}\gamma_{\nu} d)\times\\
 \nonumber
 &\times&  C_F \frac{1}{4} 
\frac{\alpha}{4\pi}\left(\frac{1}{\epsilon}+\log \frac{\mu^2}{-p^2}\right) 
 \end{eqnarray}
where $\Gamma^{NSM}\neq\gamma_{\mu}(1-\gamma_5)$ are the Lorentz
structures of the hadronic currents of the operators in
Eqs.~(\ref{eq:hc3-1})-(\ref{eq:hc3-3}) and $C_{F} = (N^{2}-1)/(2N)$ is
the standard $SU(N)$ color factor.  Applying the RGE formalism
developed for $\znbb$-decay in Ref.~\cite{Gonzalez:2015ady}, based on
\cite{Buchalla:1995vs}, we find the RGE for the Wilson coefficients
\begin{equation}
\frac{d}{d \ln(\mu)}C_i(\mu)=\gamma_{ij}C_{j}(\mu),
\label{eq:RGE}
\end{equation}
where $\gamma_{ij}$ is the matrix of the anomalous dimensions of the
corresponding operators. To leading order in the $\overline{\rm
  MS}$-scheme we find
\begin{eqnarray}\label{eq:AnomalousDimensions-1}
\gamma_{ij} = \delta_{ij}\gamma_{j},  \ \ \mbox{with} \ \ \ 
\gamma_{1}= - \gamma_{2} = - 2 \gamma_{3}  = -4 C_{F}.
\end{eqnarray}
This result is the same for different chiralities $X$ of the operators
in Eqs.~~(\ref{eq:hc3-1})-(\ref{eq:hc3-3}).  The solution of
Eq.~(\ref{eq:RGE}) is
\begin{equation}
C_{i}(\mu)=U_{ij}(\mu,\Lambda)\cdot C_{j}(\Lambda) 
\end{equation}
with the diagonal evolution matrix $U_{ij}$ linking the Wilson
coefficients at a high- and low-energy scales $\Lambda$ and $\mu$,
respectively.  Following Ref.~\cite{Gonzalez:2015ady} we find its
explicit form
\begin{eqnarray}\label{eq:RGE-U}
\nonumber
U_{ij}(\mu,\Lambda)= 
\delta_{ij}\cdot U_{j}(\mu,\Lambda), \ \mbox{with} \ U_{j}(\mu,\Lambda) = \left(\frac{\alpha_s(\Lambda)}{\alpha_s(\mu)}\right)^{\frac{\gamma_{j}}{(2\beta_0)}}
\end{eqnarray}
We take into account the quark thresholds approximately in the
standard manner \cite{Buchalla:1995vs}:
\begin{eqnarray}\label{eq:Thresh}
\nonumber
U(\mu_0,\Lambda>m_t)&=&U^{(f=3)}(\mu_0,\mu_c)U^{(f=4)}(\mu_c,\mu_b)\times\\
&\times&U^{(f=5)}(\mu_b,\mu_t)U^{(f=6)}(\mu_t,\Lambda) ,
\end{eqnarray}
with $f$ being the number of the active quarks above the threshold
$\mu_{q}$.  For $\Lambda_0=1$ TeV and $\mu_0=1$ GeV, we find:
\begin{equation}
U_{1}(\mu_0,\Lambda_0) \simeq 1.60, \ 
%
U_{2}(\mu_0,\Lambda_0) \simeq 0.6, \ 
%
U_{3}(\mu_0,\Lambda_0) \simeq 0.8.
\label{eq:u3}
\end{equation}
Thus the effect of the vertex correction in Fig.~\ref{fig:long} is at
most 60\%, as expected. This is significantly less relevant than for
the case of the SRM \cite{Gonzalez:2015ady}.

For completeness we now derive upper limits on the Wilson coefficients
$C_{i}(\Lambda)$ in Eq.~(\ref{eq:lrlagrangian}) from the current
experimental bounds on $\znbb$-decay half-life by the KamLAND-Zen
\cite{KamLAND-Zen:2016pfg} and GERDA Phase-II \cite{GERDAII:2016}
experiments both at 90\% C.L.:
\begin{eqnarray}\label{eq:T0nu-1}
%
\mbox{\cite{KamLAND-Zen:2016pfg}}
&:&  T^{0\nu}_{1/2}({}^{136}{\rm Xe}) \geq 1.07\times 10^{26} \ {\rm ys}\\
\label{eq:T0nu-2} 
\mbox{\cite{GERDAII:2016}}
&:& 
T^{0\nu}_{1/2}({}^{76}{\rm Ge})\  \geq 5.2\ \, \times 10^{25} \ {\rm ys}.
\end{eqnarray}

The QCD-corrected  $\znbb$-decay half-life formula for the LRM is
\begin{eqnarray}\label{eq:T-QCD-LRM}
\left[T^{\znbb}_{1/2}\right]^{-1}= G_{0i}\left|U_{i}(\mu_{0}, \Lambda_{0})C_{i}(\Lambda_{0})\cdot  ({\rm NME})_{i}\right|^2,
\end{eqnarray}
where $G_{0i}$ and $({\rm NME})_{i}$ are the phase-space factors
\cite{DKT} and nuclear matrix elements. The latter can be found in
\cite{Pas:1999fc,Deppisch:2012nb}. Using (\ref{eq:T-QCD-LRM}) with the
experimental bounds (\ref{eq:T0nu-1}) and (\ref{eq:T0nu-2}) we find the
upper limits on $C_{i}^{X}$ displayed in Table~\ref{t:lrWC}.

\begin{table}[h]
\begin{tabular}{c|cc|cc}
\hline 
\hline
&\multicolumn{2}{c|}{Without QCD}&\multicolumn{2}{c}{With QCD}\\
&\ \ $^{76}$Ge& \ \ $^{136}$Xe&\ \ $^{76}$Ge& \ \ $^{136}$Xe\\
\hline
$C_{1}^{L}$
&$5.3\times 10^{-9}$&$3.7\times 10^{-9}$&$3.3\times 10^{-9}$&$2.3\times 10^{-9}$\\
&&&&\\
$C_{1}^{R}$
&$5.3\times 10^{-9}$&$3.7\times 10^{-9}$&$3.3\times 10^{-9}$&$2.3\times 10^{-9}$\\
&&&&\\
$C_{2}^{L}$
&$3.1\times 10^{-10}$&$2.2\times 10^{-10}$&$5.0\times 10^{-10}$&$3.5\times 10^{-10}$\\
&&&&\\
$C_{2}^{R}$
&$8.2\times 10^{-10}$&$5.7\times 10^{-10}$&$1.4\times 10^{-9}$&$9.2\times 10^{-10}$\\
&&&&\\
%
$C_{3}^{L}$
&$2.2\times 10^{-9}$&$1.5\times 10^{-9}$&$2.7\times 10^{-9}$&$1.9\times 10^{-9}$\\
&&&&\\
$C_{3}^{R}$
&$3.4\times 10^{-7}$&$2.4\times 10^{-7}$&$4.3\times 10^{-7}$&$3.0\times 10^{-7}$\\
\hline
\hline
\end{tabular}
\caption{Individual upper limits on the Wilson coefficients in
  Eq.~(\ref{eq:lrlagrangian}), with QCD and without QCD running.
}
\label{t:lrWC}
\end{table}

\section{Limits on High-scale models from LRM $\znbb$ decay}
\label{sec:long-range limits}

\begin{figure}[h]
\centering
\includegraphics[width=0.5\linewidth]{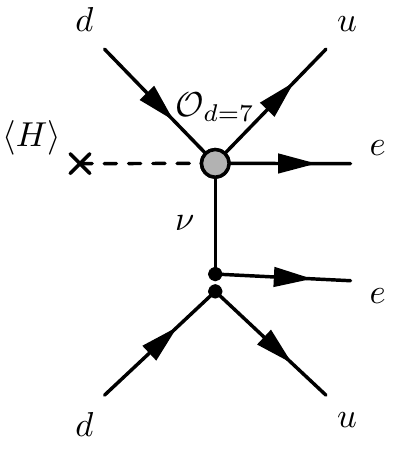}
\caption{Long-range contributions to $\znbb$ decay: From the high
  energy point of view, where $SU(2)_L\times U(1)_Y$ is unbroken,
  these contributions are generated from $d=7$ operators, always
  involving one Higgs field. At low energies, the Higgs is replaced by
  its vacuum expectation value.}
\label{fig:hslr}
\end{figure}

In order to complete our analysis we briefly discuss its impact on
high-scale models contributing to $\znbb$ via the long-range
mechanism. From the low-energy point of view, the long-range part of
$\znbb$ can be described by the Lagrangian given in
Eq. (\ref{eq:lrlagrangian}). At high-energy scales, before the
electroweak symmetry is broken, however, $\Delta L=2$ operators are
of odd dimensions. A list of all $\Delta L=2$ operators up to $d=11$ can
be found in \cite{Babu:2001ex}.  The list of $d=7$ operators
contributing to the long-range of $\znbb$ decay is the following
\cite{Helo:2016vsi,Deppisch:2015yqa}:
\begin{eqnarray}
\mathcal{O}^{d=7}_{1}&=&L^iL^jQ^kd^cH^l\epsilon_{ik}\epsilon_{jl},\\
\mathcal{O}^{d=7}_{2}&=&L^iL^jQ_i{\bar{u}}^cH^k\epsilon_{jk},\\
\mathcal{O}^{d=7}_{3}&=&L^i{\bar{e}}^c{\bar{u}}^c{\bar{d}}^cH^{j}\epsilon_{ij}.
\end{eqnarray}
For each of these operators one can form different Lorentz-invariant
contractions, corresponding to different high-scale models, see
\cite{Helo:2016vsi}. To give one example $L^iL^jQ^kd^c \to ({\bar
  L^c}L)({\bar d_R}Q)$.  At low energies, the Higgs field is replaced
by its vacuum expectation value, see Figure \ref{fig:hslr}, and we can match the Wilson
coefficients to the parameters of the high scale model via:
\begin{equation}
\frac{G_F C_{k}^{X}}{\sqrt{2}}\propto \frac{g_{eff}^3 \ v}{4\Lambda^3}.
\label{eq:lrAmcoeff}
\end{equation}
Here $\Lambda$ is the energy scale at which the $\mathcal{O}^{d=7}$
arise.  $C^{X}_k$ are the Wilson coefficients and $g_{eff}$
corresponds to an effective coupling constant, which is the geometric
mean of the three different couplings that enter in any high-scale
realization of the $C^{X}_k$.  We denoted the standard model Higgs
vev by $v$.  For $C_1^{L/R}$, for example, see Table \ref{t:lrWC},
this results in the simple estimate of a lower limit of
$\Lambda/g_{eff} \gsim 130$ $(110)$ TeV with (without) QCD
corrections.

\section{Conclusions}

We have analyzed the LNV quark-level process underlying the long-range
mechanisms of $\znbb$-decay considering its hadronization inside a
nucleus. We argued that the perturbative color-mismatched
QCD corrections, which make an appreciable impact on the theoretical
predictions in the short-range amplitude, are suppressed either by the
nuclear short-range correlation or occur at the next-to-leading order
in the long-range mechanisms. We calculated the remaining vertex
corrections and found that they do not exceed 60$\%$. We have derived
the QCD-corrected limits on the long-range mechanism Wilson
coefficients from the current experimental constraints on the
$\znbb$-decay half-life and discussed their impact on limits on
high-scale models.

\vskip5mm

\centerline{\bf Acknowledgements}

\medskip

Marcela and Carolina are grateful for the hospitality of the AHEP group in the IFIC during their visits in May-July 2016. This work was supported by the Spanish MICINN grants FPA2014-58183-P, No. SEV-2014-0398
and Multidark CSD2009-00064 (MINECO), and PROMETEOII/2014/084
(Generalitat Valenciana), and by Fondecyt (Chile) under grants No. 3150472, 
No. 1150792 and No. 3160642 as well as \mbox{CONICYT} (Chile) Ring ACT
1406 and Basal FB0821.

\setcounter{section}{0}
\def\theequation{\Alph{section}.\arabic{equation}}
\setcounter{equation}{0}


\begin{widetext}
\newpage

\begin{center}
\Large\textbf{{Erratum: QCD corrections and long-range mechanisms of neutrinoless
  double beta decay}
}
\end{center}

We have found a mistake in the renormalization prescription for the
quark fields we used in our paper \cite{Arbelaez:2016zlt}.  As a
consequence, an incorrect expression for the anomalous dimensions were
derived and shown in Eq.~(16) of Ref.~\cite{Arbelaez:2016zlt}.
The correct result to replace Eq.~(16) with is
%
%
%
%



\begin{eqnarray*}\label{eq:AnomalousDimensions-1}
\gamma_{ij} = \delta_{ij}\gamma_{j},  \ \ \mbox{with} \ \ \ 
\gamma_{1}= -3 \gamma_{2}= -6 C_{F}, \ \gamma_{3}=0.
\end{eqnarray*}
%
These changes modify the numerical values in Eq. (19) to
\begin{eqnarray*}
U_{1}(\mu_0,\Lambda_0) \simeq 2.0, \ 
%
U_{2}(\mu_0,\Lambda_0) \simeq 0.8, \ 
%
U_{3}(\mu_0,\Lambda_0) = 1
\label{eq:u3}
\end{eqnarray*}
and 
the limits in the second column \textit{"With QCD"} 
in Table I.
Here we show an updated Table I taking into account the modifications in Eq.~(19). 
As seen the numerical changes both in (19) and, as a result, in Table I are very moderate, 
about 20\%-30\%, which do not alter our conclusion about insignificance of the QCD corrections to the long-range mechanism of neutrinoless double beta decay.

\begin{table}[h]
\begin{tabular}{c|cc|cc}
\hline 
\hline
&\multicolumn{2}{c|}{Without QCD}&\multicolumn{2}{c}{With QCD}\\
&\ \ $^{76}$Ge& \ \ $^{136}$Xe&\ \ $^{76}$Ge& \ \ $^{136}$Xe\\
\hline
$C_{1}^{L}$
&$5.3\times 10^{-9}$&$3.7\times 10^{-9}$&$2.7\times 10^{-9}$&$1.9\times 10^{-9}$\\
&&&&\\
$C_{1}^{R}$
&$5.3\times 10^{-9}$&$3.7\times 10^{-9}$&$2.7\times 10^{-9}$&$1.9\times 10^{-9}$\\
&&&&\\
$C_{2}^{L}$
&$3.1\times 10^{-10}$&$2.2\times 10^{-10}$&$3.9\times 10^{-10}$&$2.8\times 10^{-10}$\\
&&&&\\
$C_{2}^{R}$
&$8.2\times 10^{-10}$&$5.7\times 10^{-10}$&$1.1\times 10^{-9}$&$7.3\times 10^{-10}$\\
&&&&\\
%
$C_{3}^{L}$
&$2.2\times 10^{-9}$&$1.5\times 10^{-9}$&$2.2\times 10^{-9}$&$1.5\times 10^{-9}$\\
&&&&\\
$C_{3}^{R}$
&$3.4\times 10^{-7}$&$2.4\times 10^{-7}$&$3.4\times 10^{-7}$&$2.4\times 10^{-7}$\\
\hline
\hline
\end{tabular}
\label{t:lrWC}
\end{table}

\ \\
\end{widetext}

\end{document}